\documentclass{article}
\usepackage[left=2.5cm, right=2.5cm, top=2cm]{geometry}
\usepackage{times}
\usepackage{w-thm}

\usepackage[]{graphicx}

\usepackage{amsmath}
\usepackage{amssymb}
\usepackage{bm}
\usepackage{amsthm}
\usepackage{import}
\usepackage{color}
\usepackage{transparent}
\usepackage{epstopdf}
\usepackage{pgfplots}
\usepackage{etoolbox}
\usepackage{subfigure}
\usepackage{xparse}
\usepackage{mathtools}
\usepackage{dsfont,bm}
\usepackage{changes}
\usepackage{siunitx}
\usepackage{eurosym}

\usepackage{changes}

\newcommand\E{\protect\rule[1.6pt]{4pt}{0.6pt}}

\newcommand{\Pd}[2]{\displaystyle \frac{\partial #1}{\partial #2} }
\newcommand{\pd}[2]{\displaystyle \tfrac{\partial #1}{\partial #2} }

\newcommand{\slfrac}[2]{\left.#1\middle/#2\right.}

\newcommand{\beq}{\begin{equation}}
\newcommand{\eeq}{\end{equation}}
\newcommand{\bal}{\begin{align}}
\newcommand{\eeal}{\end{align}}
\newcommand{\baln}{\begin{align*}}
\newcommand{\ealn}{\end{align*}}

\renewcommand \v {\bm} 
\newcommand{\mtrx}{\mathsf} 
\newcommand{\R}{\mathbb{R}} 

\newcommand{\tp}{\intercal} 
\DeclarePairedDelimiter\abs{\lvert}{\rvert}%

\newcommand{\eref}[1]{(\ref{#1})}
\newcommand{\sref}[1]{section~\ref{#1}}
\newcommand{\cref}[1]{chapter~\ref{#1}}
\newcommand{\fref}[1]{Figure~\ref{#1}}
\newcommand{\tref}[1]{Table~\ref{#1}}

\newcommand{\Fref}[1]{Figure~\ref{#1}}

\newcommand{\cDiscrete}[1]{\underline{c}_{#1}}     
\newcommand{\dDiscrete}[1]{\underline{d}_{#1}}     
\newcommand{\chengTerm}{\Theta}     

\newcommand{\fa}{f_{1}}
\newcommand{\fb}{f_{2}}
\newcommand{\fg}{f_{3}}
\newcommand{\I}[1][]{\mathcal{J}_{#1}}           
\newcommand{\f}[1][]{\widetilde{\mathcal{F}}}           
\newcommand{\fq}[1][]{\widetilde{\mathcal{F}}_{\q}}           
\newcommand{\fcv}[1][]{\widetilde{\mathcal{F}}_{\cv}}           
\newcommand{\fS}[1][]{f}           
\newcommand{\fSq}[1][]{f_{\q}}           

\newcommand{\La}{\mathcal{L}}           

\newcommand{\Ir}[1][]{\hat{\mathcal{J}}_{#1}}
\newcommand{\Jr}[1][]{\hat{\bm{\mathcal{I}}}_{#1}}                                  
\newcommand{\J}[1][]{{\bm{\mathcal{I}}}_{#1}}                                  

\newcommand{\Lar}[1][]{\hat{\mathcal{\La}}}

\newcommand{\C}{C}
\newcommand{\Cd}{\mathcal{C}}

\newcommand{\pk}{\v{p}_l}                  

\newcommand{\Ompe}{\Omega_{\textrm{pe}}}  
\newcommand{\IOmpe}[1][]{\mathds{1}_{\Ompe}} 
\newcommand{\Omc}{\Omega_{\textrm{c},{i}}}  
\newcommand{\IOmc}[1][]{\mathds{1}_{\Omc}} 
\newcommand{\cv}[1][]{{\bm \varphi}_{#1}}
\newcommand{\cvp}[1][]{{\bm{\bar{\varphi}}}_{#1}} 
\newcommand{\proj}{\mathcal{P}} 
\newcommand{\eproj}{\mathcal{P}_\textrm{ext}} 

\newcommand{\h}[1][]{{\boldsymbol{h}}_{#1}} 
\newcommand{\hr}[1][]{\hat{\boldsymbol{h}}_{#1}} 
\newcommand{\hks}{\hat{{h}}_{ks}} 

\newcommand{\cvad}{\Phi_\textrm{ad}} 

\renewcommand{\c}[1][]{\boldsymbol{c}_{#1}}                       

\newcommand{\icol}[1]{
	\left(\begin{smallmatrix}#1\end{smallmatrix}\right)%
}
\newcommand{\diag}[1]{\textrm{diag}\!\left(#1\right)}
\newcommand{\sign}[1]{\textrm{sgn}\!\left(#1\right)}

\newcommand{\cd}{d}    
\newcommand{\cdp}{\bar{d}}    

\newcommand{\q}[1][]{{\boldsymbol x}_{#1}}     
\newcommand{\qo}[1][]{ \bar{{\boldsymbol x}}_{#1}}               
\newcommand{\qA}[1][]{{\boldsymbol x}_{#1}^{*}}               
\newcommand{\qAo}[1][]{{\bar{\boldsymbol x}}_{#1}^{*}}               
\newcommand{\y}[1][]{{\boldsymbol y}_{#1}}               
\newcommand{\yo}[1][]{{\bar{\boldsymbol y}}_{#1}}               
\newcommand{\yA}[1][]{{\boldsymbol y^*}_{#1}}               

\newcommand{\z}[1][]{{\boldsymbol \theta}_{#1}}               
\newcommand{\zo}[1][]{{\bar{\boldsymbol \theta}}_{#1}}               
\newcommand{\zA}[1][]{{\boldsymbol \theta^*}_{#1}}               

\newcommand{\T}{\v{\theta}}                 
\newcommand{\qv}[1][]{{\boldsymbol q}_{#1}}                 
\newcommand{\p}{\v{p}}                 

\newcommand{\Q}{\v{Q}}                                       
\renewcommand{\H}{\v{H}} 
\renewcommand{\E}{\v{E}} 

\newcommand{\Qda}{Q_{\mathrm{d},a}}

\begin{document}



\title{An Adjoint Optimization Approach for the Topological Design of Large-Scale District Heating Networks based on Nonlinear Models}


\author{M. Blommaert$^{*,1,2,3}$, Y. Wack$^{1,2,3}$, M. Baelmans$^{1,2}$}
\date{	PUBLISHED IN: APPLIED ENERGY, Volume 280, 15 December 2020, 116025\\ DOI:10.1016/j.apenergy.2020.116025\\
	$^1$Department of Mechanical Engineering, KU Leuven, Celestijnenlaan 300 box 2421, 3001 Leuven, Belgium\\
	$^2$EnergyVille, Thor Park, Poort Genk 8310, 3600 Genk, Belgium\\
    $^3$Flemish Institute for Technological Research (VITO), Boeretang 200, 2400 Mol, Belgium\\
$^*$corresponding author, e-mail: maarten.blommaert@kuleuven.be\\}







%

\maketitle
\begin{abstract}
This article deals with the problem of finding the best topology, pipe diameter choices, and operation parameters for realistic district heating networks. Present design tools that employ non-linear flow and heat transport models for topological design are limited to small heating networks with up to 20 potential consumers. We introduce an alternative adjoint-based numerical optimization strategy to enable large-scale nonlinear thermal network optimization. In order to avoid a strong computational cost scaling with the network size, we aggregate consumer constraints with a constraint aggregation strategy. Moreover, to align this continuous optimization strategy with the discrete nature of topology optimization and pipe size choices, we present a numerical continuation strategy that gradually forces the design variables towards discrete design choices. As such, optimal network topology and pipe sizes are determined simultaneously. Finally, we demonstrate the scalability of the algorithm by designing a fictitious district heating network with 160 consumers. As a proof-of-concept, the network is optimized for minimal investment cost and pumping power, while keeping the heat supplied to the consumers within a thermal comfort range of $5 \%$. Starting from a uniform distribution of $\SI{15}{cm}$ wide piping throughout the network, the novel algorithm finds a network lay-out that reduces piping investment by $23 \%$ and pump-related costs by a factor of 14 in less than an hour on a standard laptop. Moreover, the importance of embedding the non-linear transport model is clear from a temperature-induced variation in the consumer flow rates of $72 \%$.

\end{abstract}



\setcounter{footnote}{0}
\section{Introduction}
The decarbonization of heating supplies is a crucial step in the fight against climate change. With \emph{district heating} systems, large shares of presently untapped waste heat could be harnessed for use in space heating \cite{Persson2014}. By additionally integrating renewable heating sources---such as biomass, geothermal, and solar heat---district heating systems provide a viable solution to reduce the global dependence of heating and cooling supplies on fossil fuels \cite{Werner2017}. \emph{4th generation district heating networks} that aim at fully exploiting this district heating potential while minimizing grid losses should hereby be targeted \cite{Lund2014}. The low network operation temperatures of 4th generation networks, however, put a lot of pressure on the (re-)design of district heating networks.

Recently, a number of researchers have explored the use of numerical optimization to obtain optimal network \emph{topologies}, dimensioning, and operation parameters. The discrete nature of many design variables naturally leads to the use of \emph{Mixed-Integer Programming} (MIP). In combination with the nonlinearity of the underlying district heating network physics this poses a challenging optimization problem. Given the difficult nature of such problems, many authors simplify the underlying model to obtain a Mixed-Integer Linear Program (MILP) for the optimal network. S\"oderman \cite{Soederman2007}, for example, used a MILP approach for the optimal structure and configuration of a district cooling network including cold media storage. Later, Dorfner and Hamacher \cite{Dorfner2014} used a similar approach to optimize the topology and pipe sizes of a district heating network. The application of MILP was later expanded to simultaneously optimize an increasing amount of parameters of these networks. Haikarainen et al. \cite{Haikarainen2014} simultaneously optimized the topology and operations, considering different supplier technologies and heat storage facilities. Mazairac et al. considered topology optimization of multi-carrier networks including gas and electricity \cite{Mazairac2015}. Morvay et al. \cite{Morvaj2016} combined optimal design of the network with designing the supplied energy mix. In a study to identify an optimal set of consumers to connect to a district heating network, Bordin et al. \cite{Bordin2016} optimized the topology using a MILP approach. These tools can, however, not accurately capture some important intrinsically nonlinear features of 4th generation networks. Their linearized nature fails to accurately describe the flow pattern in networks with multiple producers or with loops, the heat losses throughout the networks, or the temperature-dependent consumer heat transfer effectiveness.

More recent papers therefore shift towards at least partially capturing the nonlinearities in the district heating network models. Given the difficulty of solving a Mixed-Integer Nonlinear Program (MINLP), one popular approach is to employ heuristic optimization approaches. Tan et al. \cite{tan2018modeling} used a whale optimization metaheuristic to optimize the daily operation of a distributed energy system, while Ayele et al. \cite{ayele2019exergy} employed a particle swarm algorithm to optimize temperature levels and the size of heat pumps within a district heating network. The most commonly used heuristic approaches are evolutionary algorithms. Wang et al. \cite{wang2018modeling} optimized the pressure head of circulation pumps within a multi-source district heating network using a genetic algorithm. Their study includes a detailed nonlinear hydraulic model of the pipe network, heat sources, and substations. Hirsch et al. \cite{hirsch2018decision} performed an optimal design study of a single heat connection using a genetic algorithm. The authors optimized the design of a long distance heat transportation system, accounting for a detailed thermal and hydraulic model of the connecting pipeline. In another approach proposed by Vesterlund et al. \cite{Vesterlund2017}, a set of producer parameters in a multi-source network is optimized using a hybrid Genetic Algorithm - MILP strategy. This two layer approach uses a full thermo-hydraulic network model to optimize the producer parameters for minimal fuel costs in a nested MILP layer. None of the aforementioned heuristic approaches include the network topology in the optimization. Li and Svendsen \cite{Li2013}, conversely, used genetic algorithms for the optimization of network topology and pipe diameter dimensioning. In their work, they optimized a network of 10 consumers, while accounting for nonlinear thermo-hydraulic network modelling aspects. Discrete pipe diameters were also obtained by simple rounding.

Another strategy to cope with the nonlinearities in the network is to solve the MINLP using commercial MINLP solvers. This approach is particularly popular for solving scheduling and operational planning problems of district heating networks. Examples can be found in Deng et al. \cite{deng2017minlp} who employ it for a nonlinear scheduling optimization, and Zheng et al. \cite{zheng2018minlp} who optimize the operational planning of different heat sources in an integrated energy system. Merkert et al. \cite{merkert2019optimization} optimized the optimal operation of district heating grids while exploring globalization of the unit commitment problem. Only few authors apply MINLP solvers to network topology and pipe design problems. Marty et al. applied the DICOPT-GAMS MINLP solver to the simultaneous optimal design of the Rankine Cycle of a geothermal plant and a small heating network topology with up to 8 consumers \cite{Marty2018}, however, using a simplified network model. The same MINLP solver was also applied by Mertz et al. to the topological design and pipe sizing for an academic district heating network study case \cite{Mertz2016} and later to the design of a small existing network with 19 consumers \cite{Mertz2017}. While only Li and Svendsen \cite{Li2013}, Marty et al. \cite{Marty2018}, and Mertz et al. \cite{Mertz2016} deal with the nonlinearities that follow from the treatment of momentum and energy conservation in the topology optimization problem, the amount of design variables that can be dealt with is strongly limited by their steep computational cost scaling. This is also evident from the relatively small design problems that are considered in these papers.

The problem of computational cost scaling is further explored in a topology analysis by Allen et al. \cite{allen2020evaluation} and von Rhein et al. \cite{von2019development}. The authors investigate the steep scaling of the topological search space of a 5th generation District Heating and Cooling system for an exhaustive search. To circumvent this, they propose a minimal spanning tree heuristic to only explore a subset of the search space. The number of minimal spanning trees to explore for each possible subset of connected buildings still becomes intractable for districts of increasing size \cite{allen2020evaluation}. The design examples shown in these papers are limited to networks with four potential consumers. The argument of intractable cost scaling is further supported by the work of Weinand et al. \cite{weinand2019developing}. The authors developed a heuristic to design district heating system topologies, circumventing the steep cost scaling. Their comparative optimization approach with CPLEX was not able to exceed 7 settlements.

In contrast, optimal design applications for inherently nonlinear problems in amongst others computational fluid dynamics (CFD)---such as shape optimization of airfoils \cite{Jameson1998}, topology optimization for flow  \cite{Borrvall2003} and heat transfer  \cite{Bruns2007} problems, and even the heat exhaust optimization in nuclear fusion reactors \cite{Baelmans2017}---exploit efficient \emph{adjoint}-based optimization techniques. These methods are continuous optimization methods based on adjoint gradient calculations, which exhibit a computational cost scaling independent of the number of design variables in the optimization problem. Also in the design of flow networks, the use of these methods has been explored for topology optimization \cite{Klarbring2003}, and combined shape and topological optimization \cite{Evgrafov2006}. And most recently, in the context of district heating networks, Pizzolato et al. \cite{Pizzolato2018} applied an adjoint-based procedure to the robust hydraulic optimization of the topology, not considering thermal aspects. Around the same time, the authors of this paper published a brief first study of an adjoint-based topology optimization approach for district heating networks, emphasizing the importance of combined thermal-hydraulic modelling \cite{Blommaert2018}. Both papers relaxed the topology optimization problem to that of a continuous pipe sizing problem. Moreover, neither of these papers consider thermo-economic optimal design of district heating networks.

The contribution of this paper is to introduce a scalable adjoint-based approach for thermo-economical optimal design of district heating network topologies that simultaneously optimizes discrete pipe size and operational parameters. In contrast to simplified linearized approaches used in MILP optimization, it is first of all (1) based on a complete transport model to deal with the nonlinear effects that characterize 4th generation district heating networks. By being based on the adjoint approach, it second (2) offers good scalability towards large numbers of design variables. The approach therefore is suitable for large scale district heating network design. Third (3), it improves upon the methodology introduced by the authors of this paper \cite{Blommaert2018} by handling the discrete nature of design choices in the district heating network topology and integrating additional design constraints using aggregation techniques. Since this is a proof of concept study, we solve an optimization problem that aims at minimizing the cost associated with the installed network piping, and installed pump capacity and its operation, while meeting the thermal demand of all consumers within a reasonable tolerance. While this is not a full detailed thermo-economic cost function yet, it contains the CAPEX and OPEX contributions of a return-on-investment analysis that challenge the optimization approach most.

Several difficulties have to be overcome within this paper before the adjoint method can be considered a valid alternative to the above-mentioned MIP methods. First of all, the adjoint optimization method is continuous by nature. Since the design of topologies is discrete (to put or not to put a pipe somewhere), the method needs to be modified to handle discrete design variables. A related challenge here is how to account for costs intrinsically related to topological choices, such as the cost of the piping works. Secondly, although the adjoint method does not scale with the number of design variables, it does scale directly with the number of so-called `state' constraints. These are constraints on flow, pressure, or temperature variables that depend on the solution of a network simulation themselves. And lastly, since it is a gradient-based optimization method, it is likely that the optimization result is a local and not a global optimum. The question therefore arises whether a useful solution can be found at all.

To deal with the state constraints on the thermal energy delivery, we propose the use of constraint aggregation \cite{Kennedy2015}, in order to maintain a computational cost scaling that is independent of the number of consumers in the problem.  In order to avoid the need of relaxing the discrete optimization problem to a diameter sizing problem, a new method is introduced inspired by the smoothed Heaviside projection method from structural topology optimization \cite{Guest2004}, along with a penalization approach to account for pipe investment costs. This allows for the pipes to be selected from a discrete set of available sizes, next to the selection of the optimal topology.

The paper is outlined as follows. In \sref{sec:model}, we present a model based on conservation of mass, momentum, and energy, that governs the nonlinearities in the energy transport, and that includes a flow friction correlation valid for all relevant pipe flow regimes. Next, in \sref{sec:methods}, we introduce the optimization problem and its solution approach, including the constraint aggregation and the new projection method. Finally, in \sref{sec:results}, we show the correct functioning of the procedure on a thermal network design problem of realistic size, demonstrating its scalability. Because of the novelty of the adjoint approach for the thermo-hydraulic optimization of district heating networks, the efficient implementation of the adjoint sensitivity calculation is also elaborated in \ref{sec:A3}. In addition, the model linearizations that are needed for the adjoint approach can be exploited to build an efficient Newton solver for this highly nonlinear model. This approach is detailed in \ref{sec:A2}. The notations that are necessary to describe both the state and adjoint solvers are introduced in \ref{sec:A1}.

\section{A transport-based thermal network model}\label{sec:model}
In this paper, we will start from a physics-based thermal network model, building on conservation of mass, momentum, and energy as summarized in a previous work by the authors of this paper \cite{Blommaert2018}. In the next section the optimization approach will then be built upon this state-of-the-art model. The network will be designed for a worst-case situation, so that a steady-state model can be adopted. In this section, we will give an overview of the complete set of model equations for each network component. For clarity, we start this section by introducing the notations needed to represent the thermal network as a mathematical \emph{graph}.

\subsection{Graph representation}
\begin{figure}
	\centering
	\includegraphics[width=1.1\columnwidth]{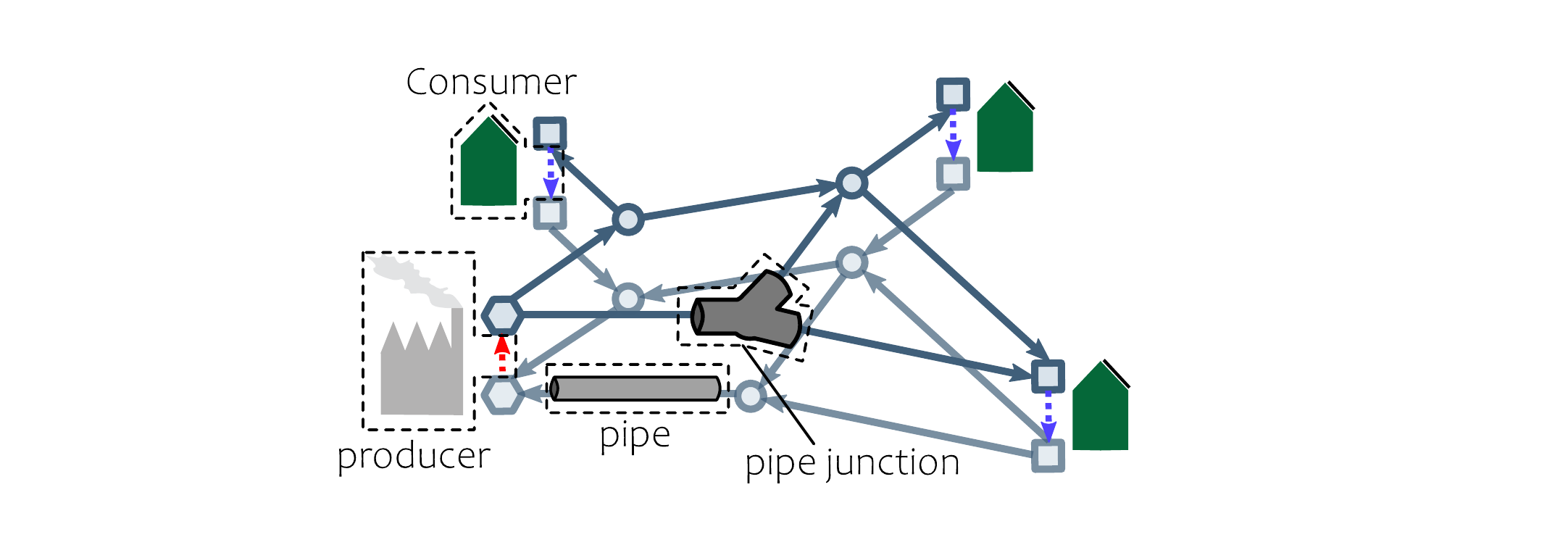}
	\caption[Slab]{ An illustration of a directed graph representation of a district heating network. Producer, consumer, and internal nodes are denoted by hexagons, squares, and circles, respectively. Dark full arrows represent feed arcs, while light full arrows represent return arcs. Dotted arrows represent producer (red) and consumer arcs (blue).
	}%
	\label{fig:network}%
\end{figure}

Referring to \fref{fig:network}, the thermal network---including pipes, junctions, producers, and consumers---can be represented by a \emph{directed graph} $G(N,A)$, with $N$ the set of all nodes and $A$ the set of all arcs in the graph. Physically, the graph consists of a feed and return network, with arcs of opposite directions. This graph, which includes all potential connections, is often referred to as the \emph{superstructure}. We can further subdivide the set of nodes $N$ into three subsets $N_p \cup N_c \cup N_i =  N$, denoting the producer, consumer, and internal nodes, respectively. Similarly, the set of arcs $A$ is partitioned into the subsets $A_p \cup A_c \cup A_i =  A$, denoting the producer, consumer, and internal arcs, respectively. The internal arcs hereby represent a geometrical pipe connection, while the other arcs rather represent state transitions at the producer and consumer ends, but at the same physical location. In what follows, we can now denote a certain network node as $n \in N$ and we can denote a directed arc going from node $i$ to node $j$ as $(i,j) \in A$, or succinctly as $ij \in A$ or even more compactly as $a \in A$.

\subsection{Network Component models}
We will consider 4 different categories of thermal network components that need modelling, as indicated with dashed boundary lines in \fref{fig:network}: pipes, pipe junctions, consumers, and producers. We outline their models here one by one.
\subsubsection{Pipe model}
For pipe flow, the momentum equation relates the gradient of the static pressure $p$ to the viscous friction. We consider water as the carrier fluid, with a constant density $\rho$, dynamic viscosity $\mu$, and specific heat capacity $c_p$, thus neglecting their temperature-dependence. In the example elaborated in \sref{sec:results}, the values will be taken as corresponding to a water temperature of $60^\circ\textrm{C}$. We then use the empirical Darcy-Weisbach equation to model the viscous pressure drop for incompressible flow as a function of the volumetric flow rate $q_{ij}$ through a pipe $(i,j)$ with length $L_{ij}$,   
\begin{equation} \label{eq:1}
(p_i - p_j)g_{ij} =  q_{ij},\quad \forall ij \in A_i.
\end{equation}
The flow conductance $g_{ij}$ in this expression is based on the correlation of Cheng \cite{Cheng2008} covering the entire range of flow regimes from laminar to turbulent, which we modified to behave regularly towards the limit of zero-sized diameters.   It is given by
\begin{equation}
g = |q|^{(\fa -1)(1-\fg)} \left(d-2\epsilon\right)^{2\fg} \chengTerm_{\mathrm{L}}^{\fa(1-\fg)} \chengTerm_{\mathrm{TS}}^{2\fb(1-\fa)(1-\fg)} \chengTerm_{\mathrm{TR}}^{2(1-\fa)(1-\fb)(1-\fg)} \chengTerm_\mathrm{C} \, , \nonumber
\end{equation}

\begin{equation}\label{eq:cheng}
	\begin{aligned}
		\mathrm{with} \quad
		\chengTerm_{\mathrm{L}} &= \frac{\rho}{16 d\pi\mu}\, , &
		\fa &= \frac{1}{1 + \left(\frac{Re}{Re_{\mathrm{LT}}}\right)^9}\, , \\
		\chengTerm_{\mathrm{TS}} &= 1.8 \log_{10}\left(\frac{Re}{6.8}\right)\, , &
		\fb &= \frac{1}{1+\left(\frac{Re}{320\left(\frac{d}{2\epsilon}\right)}\right)^2} \, , \\
		\chengTerm_{\mathrm{C}} &= \frac{d^5\pi^2}{8\rho L }\, , &
		\fg &= \frac{1}{1+\left(\frac{d-2\epsilon}{2\epsilon}\right)^6} \, , \\
		\chengTerm_{\mathrm{TR}} &= 2 \log_{10}\left(\frac{3.7 d}{\epsilon}\right) \, .
	\end{aligned}
\end{equation}

Here, $\epsilon$ denotes the absolute pipe roughness, $d$ the inner diameter of the pipe and $\textit{Re}$ the Reynolds number, defined as
$\textit{Re} = \slfrac{4 \rho \abs{q}}{(\pi \mu d)}$. The transition from laminar to the turbulent regime is controlled by $Re_{\mathrm{LT}} = 2720$.

Next, we consider the thermal aspects of an insulated pipe installed underground. First, we introduce $\theta = T - T_\infty$ as the temperature difference with the outside air temperature $T_\infty$. Let us consider $\theta_i$ to be this temperature difference at the node $i$, at which the flow enters the pipe $ij$, and $\theta_{ij}$ at the pipe exit. The pipe exit temperature difference $\theta_{ij}$, due to heat loss to the environment is given by
\begin{equation}\label{eq:3}
\theta_{ij} = \theta_i \exp{\left(\frac{-L_{ij}}{\rho c_{p} \abs{q_{ij}}{R_{t,ij}}}\right)}, \quad\forall ij \in A_i,
\end{equation}
with $R_{t,ij}$ the thermal resistance per unit pipe length between the energy carrier and the environment. For a pipe with an outer isolation casing diameter $d_{o,ij}$ that is assumed to be bigger than the inner diameter $d_{ij}$ by a fixed ratio, i.e. $d_{o,ij} = r d_{ij}$, the combined thermal resistance of pipe and soil per unit length is \cite{DEustachio1957}
\begin{equation}
R_{t,ij} = \frac{\ln(4 h/(rd_{ij}))}{2 \pi \lambda_g} +\frac{\ln{r}}{2 \pi \lambda_i},
\end{equation}
with $\lambda_i$ and $\lambda_g$ the thermal conductivity of the insulation and the surrounding ground, respectively, and $h$ the depth at which the pipe is located. The above relations for the pipe temperature drop are validated with experimental data in Ref. \cite{vanderHeijde2017}.

\subsubsection{Pipe junction model}
All nodes in the network represent pipe junctions. The flow in these junctions is governed by conservation of mass. For the incompressible flow under consideration, this reduces to conservation of the flow rate $q_{a}$ through the arcs $a$ connected to the junction, i.e.
\begin{equation} \label{eq:con}
\sum_{a=(i,n)\in A} q_{a}- \sum_{a=(n,j)\in A} q_{a} = 0,\quad \forall n \in N.
\end{equation}
Similarly, conservation of the convected energy will give us relations for the temperatures in the nodes. In the junction, perfect mixing of the incoming flows is supposed to obtain outgoing flows at the node temperature. Note that depending on the sign of the flow rate $q$ in the directed arcs connected to the node, the flow will either enter or leave the junction. Energy conservation can thus be formulated as
\begin{equation}\begin{aligned}
& \sum_{a=(i,n) \in A} \left(\max(q_a,0) \, \theta_a +\min(q_a,0) \,\theta_n\right)- \\& \sum_{a=(n,j) \in A} \left(\max(q_a,0)\,\theta_n +\min(q_{a},0)\, \theta_{a}\right) = 0, \quad \forall n \in N.
\end{aligned}
\end{equation}

\subsubsection{Consumer model}
\begin{figure}
	\centering
	\includegraphics[width=1.0\columnwidth]{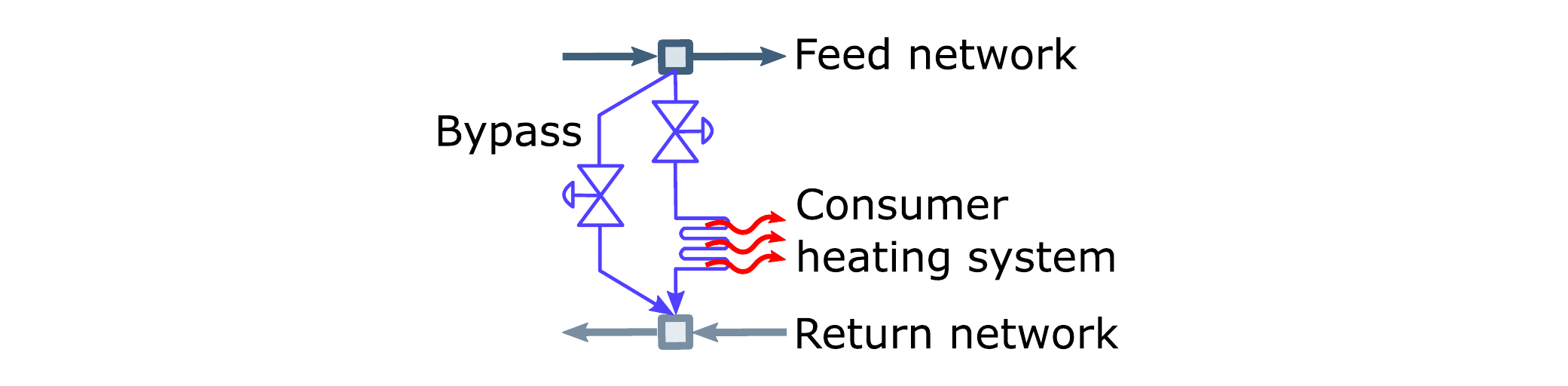}
	\caption[Consumer]{Illustration of the consumer model
	}%
	\label{fig:cons}%
\end{figure}
We now introduce a basic model to estimate the heat transferred to the consumer. We suppose the consumer substation consists of a bypass and a heating system directly connected to the network (see \fref{fig:cons}). This leads to two subsets of consumer arcs, namely the set of bypass arcs $A_{cb}\in A_c$ and the set of heating system arcs $A_{ch}\in A_c$. Both arcs have a control valve to regulate the flow. The pressure drop over the consumer heating arc is assumed to be of the form
\begin{equation} \label{eq:7}
p_i - p_j = \zeta \frac{\abs{q_{ij}}q_{ij}}{\alpha_{ij}^2}, \quad\forall ij\in A_{ch},
\end{equation}
with $\zeta$ a constant determined from nominal operating conditions \cite{pirouti2013}, and $\alpha\in \left[0,1\right]$ the valve control. It is important to note that the valve control is not trying to model the physical control of the valve, but rather attempts to obtain good conditioning for the optimization problem, while respecting the lower limit of the pressure drop ($\alpha_{ij} = 1$). Similarly, the bypass flow is controlled through the relation
\begin{equation}\label{eq:8}
q_{ij} = \beta_{ij}\frac{q_{max,b}}{\Delta p_{des,b}}\left(p_i - p_j\right), \quad\forall ij\in A_{cb},
\end{equation}
with $\beta\in \left[0,1\right]$ the bypass valve control, $q_{max,b}$ a maximal bypass flow and ${\Delta p_{des,b}}$ a typical pressure drop over the bypass.

Conservation of energy in the heating system is given by
\begin{equation}
\rho c_p q_{ij}(\theta_i - \theta_{ij}) = Q_{ij}, \quad \forall ij\in A_{ch},
\end{equation}
with $Q_{ij}$ the heat transferred to the house through the heating system.
The latter is modelled in the form of the characteristic equation for radiators \cite{ASHRAE2012,dubbel2014dubbel}:
\begin{equation}\label{eq:10}
Q_{ij} = \xi_{ij} \left(\frac{\theta_i - \theta_{ij}} {\ln{\frac{(\theta_i - \theta_{\textrm{house}})}{(\theta_{i,j} - \theta_{\textrm{house}})}}} \right)^{n_{ij}},
\end{equation}
with $\theta_{\textrm{house}}$ the known temperature difference between the indoor and environment temperature at the house. 
Values of the coefficients $\xi_{ij}$ and $n_{ij}$ are tabulated for individual radiators, according to the EN 442-2 norm \cite{EN4422}. The bypass arcs on the other hand are assumed to be free of heat losses, i.e. \begin{equation}\theta_{ij} = \theta_i \quad \forall ij\in A_{cb}\end{equation}
\subsubsection{Producer model}
In the producer arcs, a fixed input flow $q_{b}$ is imposed as boundary condition for this system of equations. In addition, a given temperature is imposed for the heat source. This leads to
\begin{equation} \label{eq:12}q_a = q_{b,a},\, \theta_a = \theta_{b,a} \quad \forall a\in A_p.\end{equation}
To uniquely define the pressures throughout the network, a reference pressure is imposed in one of the producer return nodes. Since only pressure differences influence the flow solution, the exact choice of reference does not impact the solution.

\subsection{Solution procedure}
Together with this reference pressure, equations (1-12) jointly lead to a well-defined system that, for given design and operational parameters, can be solved for the flow rates $q_{ij}$, pressures $p_i$, temperatures $\theta_i$ and $\theta_{ij}$, and consumer heat fluxes $Q_{ij}$ throughout the network. These variables are called the \emph{state variables} of the model equations and can be grouped into a single state vector $\q$. The system is solved by an efficient Newton procedure to obtain second order convergence. The analytical linearization of the model equations that is required for this also forms the basis for the adjoint solver (see \sref{sec:methods}). The structure of the model equations can be exploited to decouple the solution of flow rates and pressures at the one hand and temperatures at the other hand. We detail the procedure to efficiently solve this strongly nonlinear system of equations in \ref{sec:A2}. In order to describe the solver, we need to reformulate the equations in graph notation as an algebraic system of equations. This is done in \ref{sec:A1}.

\section{Combined size and topology optimization of thermal networks}\label{sec:methods}
For a given superstructure, we can now use the model presented in \sref{sec:model} to find an optimal configuration for a thermal network by optimizing the pipe diameters $d$, along with the network operation parameters, including the producer inflows $q_{b}$, and the consumer controls $\alpha$ and $\beta$. We will jointly denote all these \emph{design variables} with the design variable vector $\cv$.

{To formulate the design problem as a mathematical optimization problem, the first crucial step is the choice of the objective/cost function $\I(\cv,\q)$ that is to be minimized. Ideally, one would optimize the network design for the maximal return on investment or maximal energy efficiency. However, the aim of this paper is rather to demonstrate the validity of the presented approach. As such, we deliberately choose a cost function that captures only two of the main contributions, being the cost associated with the installed pump capacity and its operation, and that of the investment cost related to piping. We thus obtain the following cost function formulation that will be employed for design in a worst-case scenario:
\begin{equation}\label{eq:Ipump}
\I(\cv,\q) = \lambda_\textrm{P} \sum_{ij \in A_i}\left(p_i-p_j\right)q_{ij} + \sum_{ij \in A_i}\C(\cd_{ij})L_{ij},
\end{equation}
with $\lambda_\textrm{P}$ the price of increased pump capacity and operation, and $\C(\cd)$ a function representing the investment cost of piping placement per unit length as a function of the pipe diameter. Because of the nonlinear dependence of both terms on the pipe diameters, this choice of cost function illustrates well the advantages of the adjoint optimization strategy. Note that especially the first term is difficult to accurately represent using a linearized approximation.

Simultaneously, we impose that in this worst-case situation the heat supplied to the consumers remains within a $5 \,\%$ range of the desired heat supply $\Qda$, i.e.}
\begin{equation}\label{eq:consconstr}
0.95\,\Qda\leq Q_a\leq 1.05 \,\Qda,\quad \forall a\in A_{ch}
\end{equation}
Because these constraints depend on the state variables $\q$, they will be referred to as \emph{state constraints}. They will be succinctly denoted further using a single generic state constraint vector $\h(\cv,\q)\leq 0$. Finally, minimal and maximal values for all design variables are typically imposed, leading to \emph{box constraints} in the form $\cv[min]\leq\cv\leq\cv[max]$. These constraints are relatively easy to deal with and will be further on only implicitly denoted by an admissible set $\cvad$ of design values, giving $\cv \in \cvad$.

\subsection{Optimization problem formulation}

We can then write our design problem in the generic optimization problem formulation
\begin{equation} \label{eq:opt}\begin{aligned} \min_{\cv \in \cvad,\q}& \qquad
\I \left( \cv,\q\right) &\textrm{cost function}\\
s.t.& \qquad \c(\cv, \q) = 0, &\textrm{model equations}\\
& \qquad \h(\cv,\q) \leq 0, &\textrm{state constraints}
\end{aligned}\end{equation}
with $\c(\cv,\q) = 0$ succinctly denoting the model equations (1-12), as detailed in \ref{sec:A1}.

Since the set of model equations $\c(\cv,\q) = 0$ we defined in \sref{sec:model} can be solved for a state solution $\q(\cv)$ for each evaluated network design $\cv$, we can use them to eliminate the state variables $\q$ in the optimization problem. The so-called \emph{reduced optimization problem} then reads
\begin{equation}
\begin{aligned} \label{eq:redopt} \min_{\cv \in \cvad}& \qquad
\Ir \left( \cv\right) &\\
s.t. & \qquad \hr(\cv) \leq 0, & \textrm{state constraints},
\end{aligned}\end{equation}
with $\Ir$ and $\hr$ the reduced cost function and reduced state constraint vector, respectively. The adjoint optimization strategy discussed in the remainder of this section gives us an efficient approach to solve this optimization problem.

\subsection{Adjoint-based continuous optimization procedure}
One possible strategy to solve the nonlinear optimization problem \eqref{eq:redopt} is by using a \emph{Sequential Quadratic Programming} (SQP) strategy. One hereby linearises the constrained optimization problem in optimization iteration $l$ to a linearly-constrained quadratic program for the step size $\pk=\cv[l+1]-\cv[l]$,
\begin{equation}
\begin{aligned} \label{sqpstep}\min_{{\pk \in \Phi_p = \left\{ \pk \left|\, \cv[l] + \pk \, \in \cvad    \right\}\right. }}& \qquad {\nabla \Ir}\left( \cv[l]\right)^\tp \pk + \frac{1}{2} \pk^\tp \mtrx{B}_k \pk\\
 s.t.& \qquad \v{\widehat{h}}(\cv[l]) + \nabla \v{\widehat{h}}\left(\cv[l]\right)^\tp \pk\leq0,
\end{aligned} \end{equation}
where $\mtrx{B}_k$ is an approximation of the reduced hessian $\nabla^2 \Ir$, which will be estimated from gradient information in subsequent optimization iterations with a damped BFGS algorithm \cite{Nocedal2006}. It is clear that the main challenge is thus the efficient calculation of the cost function and state constraint sensitivities $\nabla \Ir$ and $\nabla \hr$. For this purpose, the adjoint approach is used.

Because of the novelty of the adjoint approach in its application to district heating network optimization, we will elaborate next on how to evaluate these sensitivities using an \emph{adjoint} approach.
For compactness of notation, we will leave out the operator arguments below if they are clear from the context. In addition, we define the vector function  $\J = [\I,\h^\tp]^\tp$ that groups the quantities of interest of which we need the sensitivities to solve the quadratic program \eqref{sqpstep} for an optimization step $\pk$. Its sensitivities are then given by the sensitivity matrix $\nabla \Jr := \frac{d \Jr}{d\cv}^\tp$.

Using the chain rule of differentiation, we can decompose this matrix into two contributions, i.e.
\begin{equation}\label{eq:chainrule}
\frac{d \Jr}{d\cv} = \Pd{\J}{\cv}+\Pd{\J}{\q}\frac{d\q}{d\cv}.
\end{equation}
It should be noted that while the first term is a quite simple term that only contains the direct dependence of the cost function on our design variables (zero in our case), the second term features the nonlinear dependence $\frac{d\q}{d\cv}$ of the model equation solution. An expression for $\frac{d\q}{d\cv}$ can be found by linearising the model equations $\c(\cv,\q) = 0 $ around the design variables $\cv[l]$ and model solution $\qo[l] = \q(\cv[l])$ of the current iterate, giving
\begin{equation}\label{eq:linstate}
\Pd{\c}{\q} \frac{d\q}{d\cv} = - \Pd{\c}{\cv}.
\end{equation}
It can be observed that the number of linear algebraic sets of equations that need to be solved to evaluate the complete sensitivity matrix $\frac{d\q}{d\cv}$ scales directly with the number of design variables in $\cv$. The adjoint method avoids this in an elegant way. It can be easily derived by substituting \eqref{eq:linstate} in \eqref{eq:chainrule}. After rearranging terms we get
\begin{equation*}
\frac{d \Jr}{d\cv} = \Pd{\J}{\cv}+\left[-\left(\Pd{\c}{\q}\right)^{-\tp}\left(\Pd{\J}{\q}\right)^\tp\right]^\tp\Pd{\c}{\cv}.
\end{equation*}
The term between brackets can be obtained by solving the \emph{adjoint equation}
 \begin{equation}\label{eq:adjoint}
 \left(\Pd{\c}{\q}\right)^\tp \qA=-\left(\Pd{\J}{\q}\right)^\tp,
 \end{equation}
 for the so-called \emph{adjoint variables} $\qAo[l] = \qA(\cv[l],\qo[l])$. The derivative matrix in optimization iteration $l$ then evaluates as
 \begin{equation}\label{eq:design}
 \nabla \Jr(\cv[l]) = \left(\Pd{\J}{\cv}(\cv[l],\qo[l])\right)^\tp+\left(\Pd{\c}{\cv}(\cv[l],\qo[l])\right)^\tp\qAo[l].
 \end{equation}

 This latter expression is cheap to evaluate, since $\pd{\c}{\cv}$ can be derived analytically and evaluated. The main computational cost is therefore in the solution of the adjoint equation \eqref{eq:adjoint}. It can be observed that the number of algebraic system solutions needed to solve the adjoint equation is now independent of the number of design variable entries in $\cv$, but rather scales with the number of quantities of interest in $\J$. For an unconstrained optimization problem with a scalar-valued cost function, the gradient calculation cost is therefore reduced to a single system solution. In the current optimization problem formulation, an additional system solution is needed for each state constraint.

 We therefore choose to limit the number of state constraints drastically by aggregating all these constraints using a variant of the Kreisselmeier-Steinhauser function \cite{Kennedy2015}, turning these constraints into a single scalar-valued constraint $\hks(\cv,\gamma) \leq 0$, with
 \begin{equation}
 \hks(\cv,\gamma) = \frac{1}{\gamma} \ln{\sum_{i = 1 }^{n_h} e^{\gamma [\hr]_i}}.
 \end{equation}
 In this equation, $[\hr]_i$ represents a specific component of the vector $\hr$ with length $n_h$, and $\gamma \in ]0,\infty[$ is a parameter that controls the trade-off between smoothness and numerical conditioning of the approximation on the one hand, and exactness of the approximation on the other. In practice, numerical continuation is used to increase it's value $\gamma$ in different stages of the optimization as explained in \sref{sec:numcont}. Using this aggregated constraint in combination with the adjoint method we effectively reduce the computational costs of the complete gradient matrix evaluation to approximately two algebraic system solutions, independent of the number of design variables. Thus, a scalable optimization approach is achieved.

\subsection{A numerical continuation strategy for discrete diameter optimization}

The algorithm presented thus far accounts only for continuous optimization variables, such as the pipe diameters. This could, in theory, lead to a whole range of diameters, with some of them perhaps of negligible size. In order to perform a true topology optimization we need to include the choice whether or not to put a pipe in a particular arc of the superstructure. We therefore modify the diameter design variable to either not place a pipe at all ($d \rightarrow 0$), or to choose from a number of standard pipe sizes. {We then end up with a discrete design variable $d \in \mathcal{D}$, with $\mathcal{D}$ a discrete set of diameters, e.g. $\mathcal{D} = \left\{0,0.05,0.125,0.20\right\} \,\textrm{m}$.\footnote{Note that we in practice substitute the discrete diameter ``0'' in this set with a small but non-zero diameter at which the model equations are still well-defined and their solution numerically well-conditioned.} To achieve this discrete nature without giving up the benefits of the adjoint approach, we introduce a novel numerical continuation strategy that gradually forces the continuous optimization method towards these discrete diameter choices. The proposed strategy consists of two elements: the \emph{smoothed projection of the diameters} onto the discrete diameter set and the \emph{gradual penalization of intermediate diameter values} through the piping cost relation $\C(d)$. We will now elaborate these components. }

{
\subsubsection{Smoothed multi-projection method for discrete diameter optimization}
For the projection of the diameters on the discrete diameter set $\mathcal{D}$, we modify the so-called \emph{projection method} that is used in the field of structural topology optimization to obtain binary variables with a continuous optimization method \cite{Guest2004}. }
This technique uses a smooth approximation of a Heaviside function with described by \cite{Wang2011}
\begin{equation}\label{eq:proj}
\bar{\varphi} = \proj\left(\varphi,\eta,\chi\right) = \frac{\tanh\left(\chi \eta\right) +\tanh\left(\chi (\varphi-\eta)\right)}{\tanh\left(\chi \eta\right) +\tanh\left(\chi (1-\eta)\right)}
\end{equation}
to project the design variable $\varphi$ on a binary decision variable $\bar{\varphi}\in \left\{0,1\right\}$ that determines whether or not material should be placed at a certain location in a structure. The variables $\chi\in ]0,\infty[$ and $\eta\in [0,1]$ are two factors controlling, respectively, the steepness of the Heaviside approximation and the threshold value above which the variable $\varphi$ is projected onto to upper limit.

\begin{figure}
	\centering
	\makebox[\textwidth][c]{	\includegraphics[width=\paperwidth]{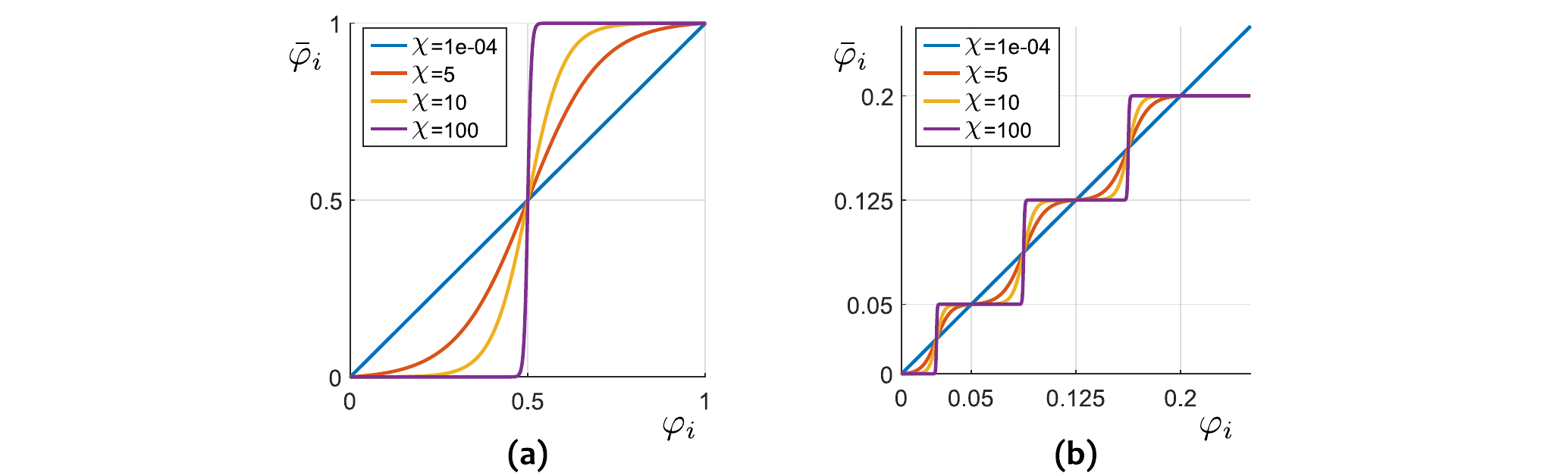}}
	\caption[projection]{ (a) Illustration of the smoothed projection of a design variable $\varphi_\textrm{i}$ onto a binary design variable $\bar{\varphi}_\textrm{i}$. (b) Illustration of the extended smoothed Heaviside projection of $\varphi_\textrm{i}$ onto a variable $\bar{\varphi}_\textrm{i}$ onto some discrete set, chosen here to be a set of pipe diameters $\mathcal{D} = \left\{0,0.05,0.125,0.20\right\} \,\textrm{m}$.
	}%
	\label{fig:proj}%
\end{figure}

We extend this principle here to allow projection on the multiple discrete variables in $\mathcal{D}$. To this end, we construct an extended projection operation $\cdp = \eproj\left(\cd,\chi\right)$ that is a piecewise function assembled from rescaled versions of the projection operation $\proj$, as shown in \fref{fig:proj}. The extended projection then uses the basic projection function $\proj\left(\cd,0.5,\chi\right)$ to project inbetween two discrete values, as well as project downwards (i.e. $\proj\left(\cd,0,\chi\right)$) for values bigger than the largest value in $\mathcal{D}$. For a diameter $\cd$ and discrete design variable set $\mathcal{D} = \left\{0, \dDiscrete{1},\dDiscrete{2},...,\dDiscrete{n} \right\}$, the extended projection operator then becomes
\begin{equation}
\cdp = \eproj\left(\cd,\chi\right) = \left\{  \begin{array}{ll}
\dDiscrete{1} \proj\left(\frac{\cd}{\dDiscrete{1}},0.5,\chi\right) & \cd\leq \dDiscrete{1}\\
\dDiscrete{j} + \left(\dDiscrete{j+1}-\dDiscrete{j}\right) \proj\left(\frac{\cd-\dDiscrete{j}}{\dDiscrete{j+1}-\dDiscrete{j}},0.5,\chi\right) & \dDiscrete{j}\leq\cd\leq \dDiscrete{j+1}\\
\dDiscrete{n} + w \proj\left(\frac{\cd-\dDiscrete{n}}{w},0,\chi\right) & \dDiscrete{n}\leq\cd\\
\end{array}, \right.
\end{equation}
with $w$ a parameter set such that $\cd< \dDiscrete{n} +w$ for all values of $\cd$ and where we assume that $\cd$ is always positive.

{
\subsubsection{Piping cost penalization}}
{
In the beginning of this section we introduced $\C(\cd)$ as a function that represents the piping cost as a function of the pipe diameter $d$. It should be noted, however, that the only relevant values of this function are those in the discrete set $\Cd =\left\{0,\cDiscrete{1},\cDiscrete{2},\dots,\cDiscrete{n}\right\}$ corresponding to the discrete diameters $d \in \mathcal{D}$, since a cost can only be provided for these available pipe sizes. For the values in between, a continuous interpolation is to be provided to account for the gradient-based optimization approach. We choose this interpolation as to deliberately make `intermediate' values ($d \not\in \mathcal{D}$) unfavorable for the optimization. Following this reasoning, we smoothly approximate a piecewise constant function that features a strong increase in price when the diameter exceeds a value in the set $\mathcal{D}$. Such a function can be described by the piecewise function}
\begin{equation}
\C(d,\upsilon,\omega) = \left\{  \begin{array}{ll}
\upsilon\cDiscrete{1} \proj\left(\frac{\cd}{\dDiscrete{1}},0,\omega\right)+\left( 1 - \upsilon\right) \frac{\cDiscrete{n}\left(d - \dDiscrete{1}\right)}{\dDiscrete{n}-\dDiscrete{1}}  & \cd\leq \dDiscrete{1}\\
\upsilon\left(\cDiscrete{j} + \left(\cDiscrete{j+1}-\cDiscrete{j}\right) \proj\left(\frac{\cd-\dDiscrete{j}}{\dDiscrete{j+1}-\dDiscrete{j}},0,\omega\right)\right)+\left( 1 - \upsilon\right) \frac{\cDiscrete{n}\left(d - \dDiscrete{1}\right)}{\dDiscrete{n}-\dDiscrete{1}} & \dDiscrete{j}\leq\cd\leq \dDiscrete{j+1}\\
\upsilon\left(\cDiscrete{n} + \cDiscrete{w} \proj\left(\frac{\cd-\dDiscrete{n}}{w},0,\omega\right)\right) +\left( 1 - \upsilon\right) \frac{\cDiscrete{n}\left(d - \dDiscrete{1}\right)}{\dDiscrete{n}-\dDiscrete{1}} & \dDiscrete{n}\leq\cd\\
\end{array}. \right.
\end{equation}

\begin{figure}
	\centering
	\makebox[\linewidth][c]{	\includegraphics[width=\linewidth]{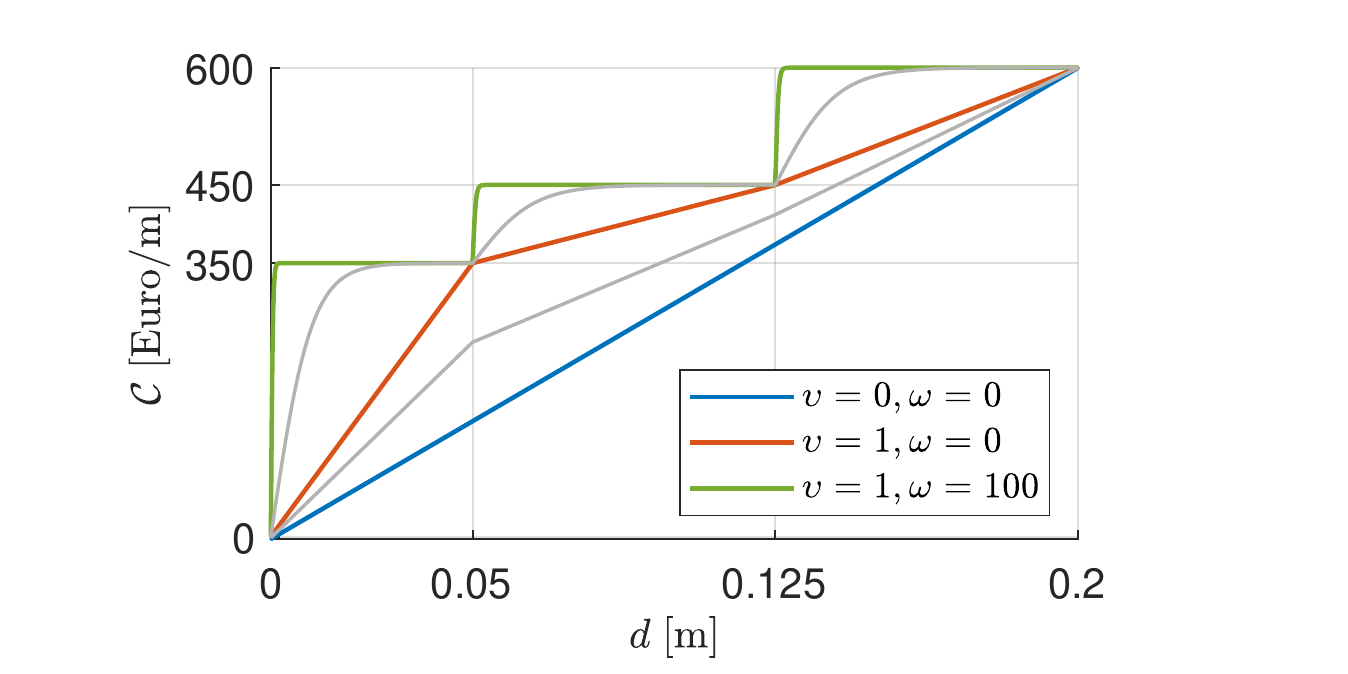}}
	\caption[investment]{Evolution of the penalization on the pipe investment cost $\mathcal{C}$. The optimization starts on a linear function (blue) that gets pushed towards a piecewise linear function (orange) by increasing $\omega$. Next the cost is pushed towards a Heaviside step function (green) by $\upsilon$, representing the real cost of piping. The grey lines illustrate intermediate steps in the continuation.}%
	\label{fig:invest}%
\end{figure}

Here, $\cDiscrete{w}$ is an arbitrary high cost value to penalize oversized diameters\footnote{Oversized diameters are prevented with an appropriate choice of the box constraints, represented by the admissible set $\cvad$.}.
The variable $\upsilon\in [0,1]$ controls the interpolation between the piecewise linear- and the piecewise constant function. $\omega\in ]0,\infty[$ controls the steepness of the Heaviside approximation for the piecewise constant function. The process of interpolation and steepness increase can be seen in \fref{fig:invest}.
{
Given the ill-posedness that the steep derivatives of this function induce on the optimization problem, a numerical continuation strategy is used to relax this problem in the initial stages of optimization. This will be explained next.
\subsubsection{Numerical continuation strategy}}
\label{sec:numcont}
After including the aggregated constraint, the diameter projection, and the continuous piping cost function $\C(\varphi_i,\upsilon,\omega)$, the optimization problem formulation reads\footnote{It is understood that the extended projection operator $\cdp = \eproj\left(\cd,\chi\right)$ can be generalized to operate on the complete vector of design variables, i.e. $\cvp = \eproj\left(\cv,\chi\right)$, in such a way that only the discrete diameter variable components of $\cv$ are affected. }
\begin{equation}\begin{aligned} \label{eq:projopt} \min_{\cv \in \cvad}& \qquad
\Ir \left( \cvp,\upsilon,\omega\right) &\\
s.t. & \qquad \hks(\cv,\gamma) \leq 0, & \textrm{aggregated constraint},\\
& \qquad \cvp = \eproj\left(\cv,\chi\right) \leq 0, & \textrm{extended projection},
\end{aligned}\end{equation}
in which the continuation variables $\upsilon$, $\omega$, $\gamma$, and $\chi$ control the trade-off between relaxing the discrete problem that is to be solved to increase the well-posedness of the optimization problem, and the exactness of its approximation. We will therefore apply numerical continuation to gradually force the design towards the discrete solution we are interested in. That is, we add an outer loop to the optimization in which the above optimization problem is solved for different values of the continuation variables, and initialize the optimization procedure with the solution of the last continuation iteration to overcome the problem of ill-conditioning.

Note that the cost function and state constraints are evaluated in \eref{eq:projopt} based on the projected design variable. Using the chain rule of differentiation, we find that the derivative of the quantities of interest is then given by
\begin{equation}
\frac{d \Jr}{d\cv} = \frac{d \Jr}{d\cvp} \Pd{\eproj}{\cv},
\end{equation}
where $\frac{d \Jr}{d\cvp}$ can be efficiently calculated by the adjoint gradient calculation \eqref{eq:design} as before and multiplied with the projection gradient.

\section{Demonstration of the optimal design approach on a fictitious design problem}\label{sec:results}
\begin{figure}
	\centering
	\makebox[\linewidth][c]{	\includegraphics[width=\linewidth]{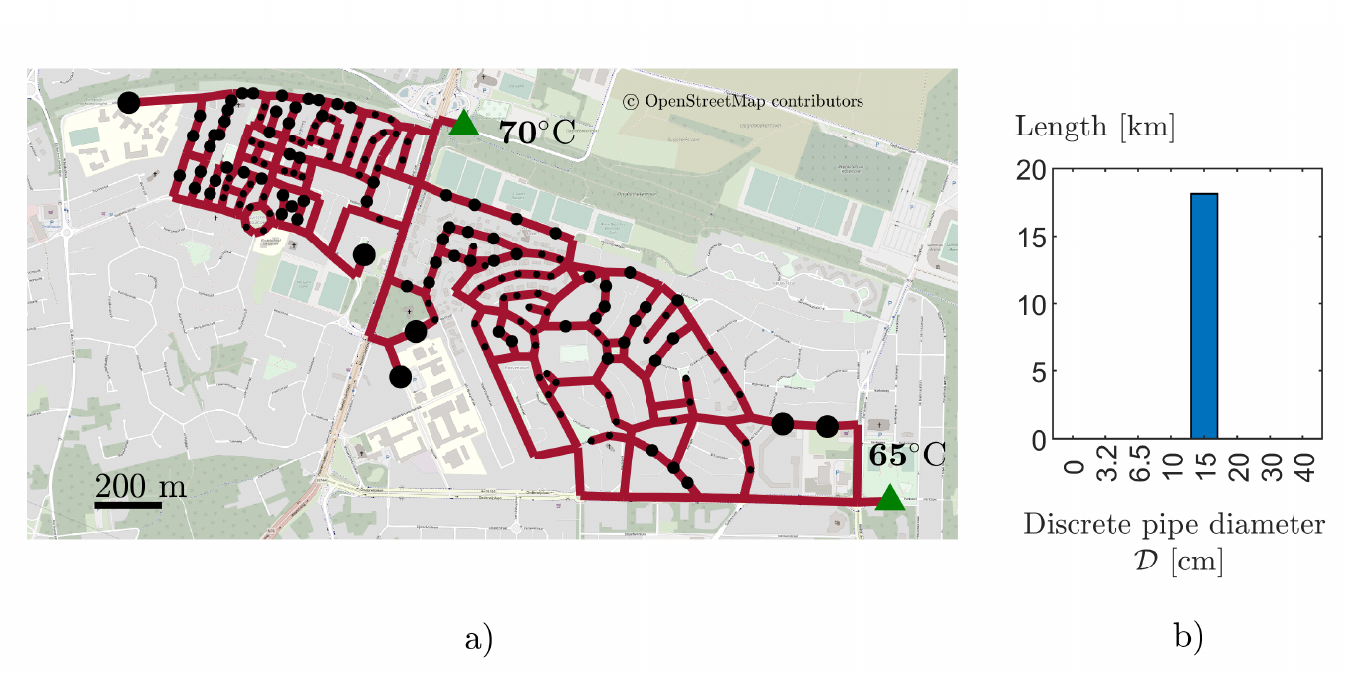}}
	\caption[projection]{a) Initial topology and configuration of the network. The green triangles represent heat producers of $T_{\mathrm{b},2} = \SI{70}{\degreeCelsius}$ (left) and $T_{\mathrm{b},1} = \SI{65}{\degreeCelsius}$ (right). Consumers are represented by black circles of varying sizes corresponding to their respective heat demand ($\SI{5}{\kilo\watt}$, $\SI{15}{\kilo\watt}$, $\SI{50}{\kilo\watt}$). b) Total length of all placed discrete pipe diameters in the feed network.}%
	\label{fig:warmstart}%
\end{figure}

In this section, we will demonstrate the functioning and scalability of the network optimization algorithm on a fictive test case.
Yet, we choose a somewhat realistic test case that aims at designing a medium-sized district in Waterschei, Genk. We take a case set-up with two producers at different inflow temperature levels ($T_{\mathrm{b},1} = \SI{65}{\degreeCelsius},\, T_{\mathrm{b},2} = \SI{70}{\degreeCelsius}$). 160 consumers of 3 different consumer types are distributed throughout the neighbourhood. A summary of the different consumer characteristics can be found in \tref{tab:consumer}. As can be seen in \Fref{fig:warmstart}, the pipe superstructure for the network optimization is placed onto a part of the neighborhood's street grid. In this test case, the total amount of design variables is 632. To the best of the authors' knowledge, it is the largest reported application of topology optimization for district heating networks that is based on a fully-fledged flow and heat transport model. 

\begin{table}[h]
	\centering
	\caption{Consumer characteristics}
	\label{tab:consumer}
	\begin{tabular}{llllllll}
	\textbf{Type} 	& $\mathbf{Q_\mathrm{d,a}}$  & $\mathbf{T_\mathrm{in,d}}$ &$\mathbf{\Delta T_\mathrm{d}}$& $\mathbf{n_\mathrm{d}}$&$\mathbf{\xi}$ $[\frac{\textrm{kW}}{\textrm{K}^{n_\mathrm{d}}}]$ & $\mathbf{\zeta}$ $[10^{12} \,\frac{\si{\pascal\second}}{\si{\cubic\meter}}]$\\
	Dwelling & $\SI{15}{\kilo\watt}$	& $\SI{55}{\degreeCelsius}$ & $\SI{20}{\degreeCelsius}$ & $1.2$ & $0.34$& $1.5$\\
	Renovated dwelling & $\SI{5}{\kilo\watt}$	 & $\SI{55}{\degreeCelsius}$ & $\SI{20}{\degreeCelsius}$ & $1.42$ & $0.06 $&$0.14$\\
	Commercial demand & $\SI{50}{\kilo\watt}$	 & $\SI{55}{\degreeCelsius}$& $\SI{20}{\degreeCelsius}$ & $1.2$ & $1.23$& $13.54$ \\
	\end{tabular}
\end{table}

Assuming a worst-case situation with an outdoor temperature $T_\infty = \SI{-8}{\degreeCelsius}$, we now try to design the network to have the least possible investment cost, as well as pump capacity and operation cost. The weight for pump-related costs was chosen to be $\lambda_{\mathrm{P}}=10^2\,\si{\textrm{M\euro}\per\kilo\watt}$ in the current cost function definition, in order to limit the pressure drops in the piping to reasonable engineering limits. The design is initialized by distributing a uniform pipe network with a diameter of $d=\SI{0.15}{m}$ over the entire superstructure (see \Fref{fig:warmstart}). Initially the control valves (heating system and bypass valves, as well as producer inflows) are fully opened ($\alpha = 1,\, \beta = 1, q_{\mathrm{b}}= q_{\mathrm{b,max}}$). {To start the optimization within the heat satisfaction boundaries set by \eqref{eq:consconstr}, we first optimize the network operation parameters for minimal consumer heat dissatisfaction, i.e.}
\begin{equation}
	\I(\cv,\q) = \left(Q_\mathrm{a}-Q_\mathrm{d,a}\right)^2 \, ,
\end{equation}
while fixing the pipe diameters. To choose reasonable consumer heating characteristics, we suppose the consumer heating systems are designed to provide this heat load at nominal/design conditions, described next. First of all, we assume the design is based on a desired temperature drop $\Delta T_{\textrm{d}}$ and pressure drop $\Delta p_{\textrm{d}}$ of $\SI{50}{\kilo\pascal}$ over the heating system. Next, we take the heating system inflow temperature $T_{\textrm{in,d}}$ and a room temperature of $T_\mathrm{room,d} = \SI{20}{\degreeCelsius}$. We can then obtain the heating system characteristics from
 \begin{equation}\begin{aligned}
\xi &\approx Q_\mathrm{d}{\left(\frac{T_{\textrm{in,d}} - T_{\textrm{out,d}}} {\ln{\frac{(T_{\textrm{in,d}} - \theta_{\textrm{house}})}{(T_{\textrm{out,d}} - \theta_{\textrm{house}})}}} \right)} ^{-n_{ij}} \quad \mathrm{with} \; T_{\textrm{out,d}} = T_{\textrm{in,d}}-\Delta T_{\textrm{d}},\\
\zeta &= \frac{\Delta p_{\textrm{d}}}{q_{\textrm{d}}^2} = \frac{\Delta p_{\textrm{d}}*(\rho c_p \Delta T_{\textrm{d}})^2}{Q_\mathrm{d}^2} ,
\end{aligned}\end{equation}
using the coefficient $n$ found in \tref{tab:consumer} as a typical value for heating systems (see Ref. \cite{ASHRAE2012}).

We suppose the producers deliver heat to the network at a constant temperature of $T_{\mathrm{b},1} = \SI{65}{\degreeCelsius}$ and $T_{\mathrm{b},2} = \SI{70}{\degreeCelsius}$. Consequently, we can directly relate the controlled volumetric inflow $q_\mathrm{b}$ at the producers to the heat inflow via
\begin{equation}
Q_{in} = \rho c_p q_b (T_b - T_{\textrm{out,d}}),
\end{equation}
with $T_{\text{out,d}} = \SI{40}{\degreeCelsius}$ the design outflow temperature of the network. We can thus directly control the maximal input heat delivered by the producers with a simple constraint on the inflow $q_b$. Given a total heat demand of $\SI{1.77}{\mega\watt}$ for the 160 consumers, we can only satisfy the demand if the combined maximal heat provided by both producers is big enough to cover both this demand and the heat losses on the feed line.

After the aforementioned warm-starting approach, we apply the optimization procedure described in \sref{sec:methods} to optimize the layout and pipe dimensions of this network, along with the producer inflow and consumer valve settings. A continuation strategy with 20 iterations was used for updating the continuation variables in the aggregated constraint, the smoothed multi-projection and the investment cost penalization, increasing them within the ranges $\gamma \in [5\cdot10^3,10^5]$, $\chi \in [0,100]$, $\upsilon \in [0,1]$ and $\omega \in [0,100]$. The maximal heat input $Q_{\textrm{in,max,1}} = Q_{\textrm{in,max,2}} =  \SI{2}{\mega\watt}$ of each individual producer was chosen high enough to supply the entire network. We consider a set of available pipe diameter choices $\mathcal{D}$ and their corresponding cost for road works and piping $\mathcal{C}$ as tabulated in \tref{tab:discreteValues}.

\begin{figure}
	\centering
	\makebox[\linewidth][c]{	\includegraphics[width=\linewidth]{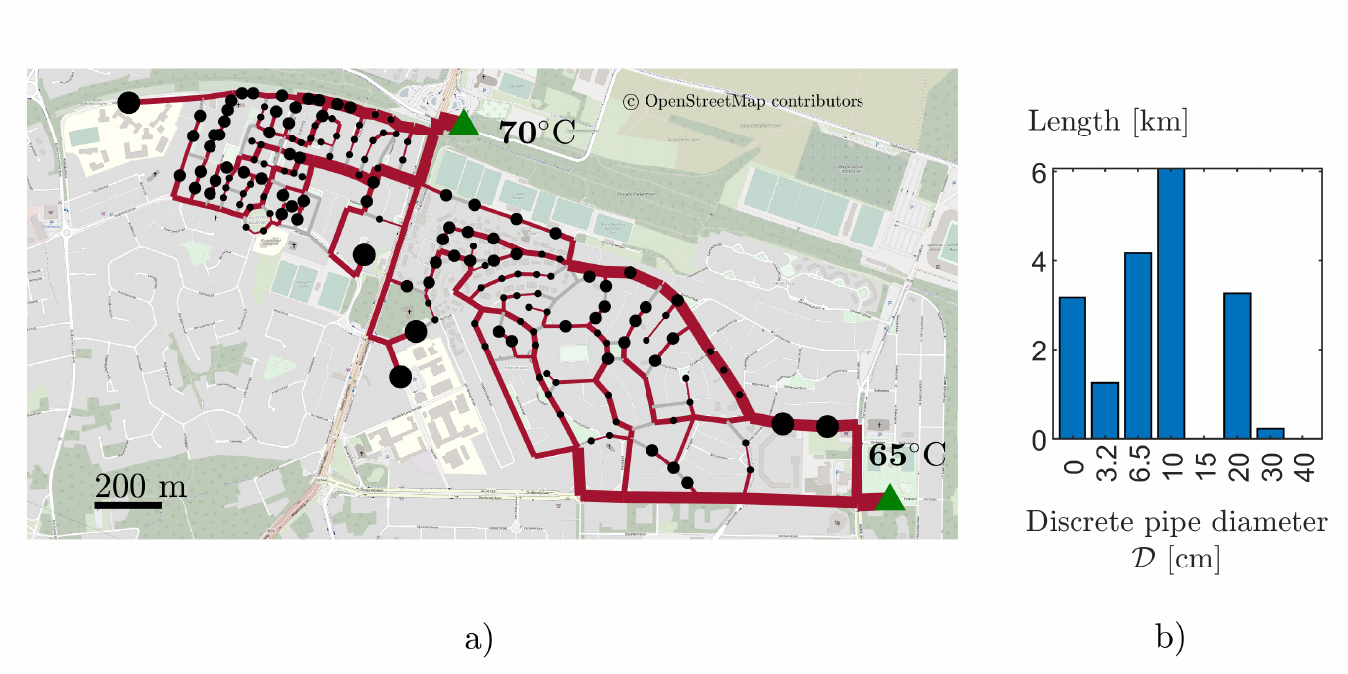}}
	\caption[projection]{a) Optimal network topology and configuration. The green triangles represent heat producers of $T_{\mathrm{b},2} = \SI{70}{\degreeCelsius}$ (left) and $T_{\mathrm{b},1} = \SI{65}{\degreeCelsius}$ (right). Consumers are represented by black circles of varying sizes corresponding to their respective heat demand ($\SI{5}{\kilo\watt}$, $\SI{15}{\kilo\watt}$, $\SI{50}{\kilo\watt}$). Placed pipes are drawn as red lines. Removed pipes are drawn in grey. The resulting topology shows a clear separation of the two neighborhoods. b) Total length of all placed discrete pipe diameters in the feed network.}%
	\label{fig:topology}%
\end{figure}

\begin{table} [h]
	\caption{Discrete diameters $\mathcal{D}$ available and their respective investment cost $\Cd$}
	\label{tab:discreteValues}
	\centering
	\begin{tabular}{l|llllllll}
		$\mathcal{D} \; [\si{\meter}]$ & 0 & 0.032 & 0.065 & 0.1 & 0.15 & 0.2 & 0.3 & 0.4 \\
		$\Cd \; [\si{\text{\euro}\per\meter}]$ & 0 & 2202  & 2218  & 2258& 2448 & 2461&2665 &2922
	\end{tabular}
\end{table}
 For this test case with 632 design variables, the optimization procedure, including warm-start, converged after $\SI{46}{\minute}$. The optimization was run on a single CPU of a standard laptop with an Intel Core i7 processor (1.9 GHz). A decrease in the piping investment cost by $\SI{23}{\percent}$ from $\SI{88.8}{\mega\text{\euro}} $ to $\SI{68.7}{\mega\text{\euro}}$ with respect to the uniform pipe distribution design of \Fref{fig:warmstart} was achieved (i.e. the design obtained after exclusively optimizing the network operation variables). It is understood that this uniform design features $\SI{0.15}{m}$ pipes throughout the superstructure and that the optimized operation parameters satisfy the consumer heat demand up to a very high tolerance. The pump operation cost was reduced by a factor of 14 with respect to this `uniform' pipe design. The resulting network topology is given in \fref{fig:topology}a. It can be directly observed that larger pipe diameters are provided for high demand consumers, due to their greater flow needs.  Moreover, the resulting pipe diameters all have discrete values thanks to the smoothed multi-projection strategy presented in this paper. The total length of pipe installed for each of the different pipe size options can be seen in \fref{fig:topology}b. It should be noted that the topology optimization algorithm explicitly decided not to install any piping in about $\SI{3}{\kilo\meter}$ of the superstructure.

The heat demands of all consumers are met with a producer inflow $Q_{\textrm{in},1}= \SI{0.92}{\mega\watt}$ and $Q_{\textrm{in},2} = \SI{1.09}{\mega\watt}$. Minimizing the pump operational cost \eqref{eq:Ipump} the algorithm is able to push the supplied consumer heat $Q_{\mathrm{a}}$ close to the lower limit of the thermal comfort range of the consumer \eqref{eq:consconstr}. It is intuitive that the cost of the network design will indeed be lowest if these constraints are only marginally satisfied. Indeed, through the increase of the constraint smoothness parameter $\gamma$ in the numerical continuation strategy, the algorithm achieves a maximum deviation of the consumer heat loads from the lower limit of $0.008 \,\%$ while conservatively satisfying \eqref{eq:consconstr}.

\begin{figure}
	\centering
	\makebox[\linewidth][c]{	\includegraphics[width=\linewidth]{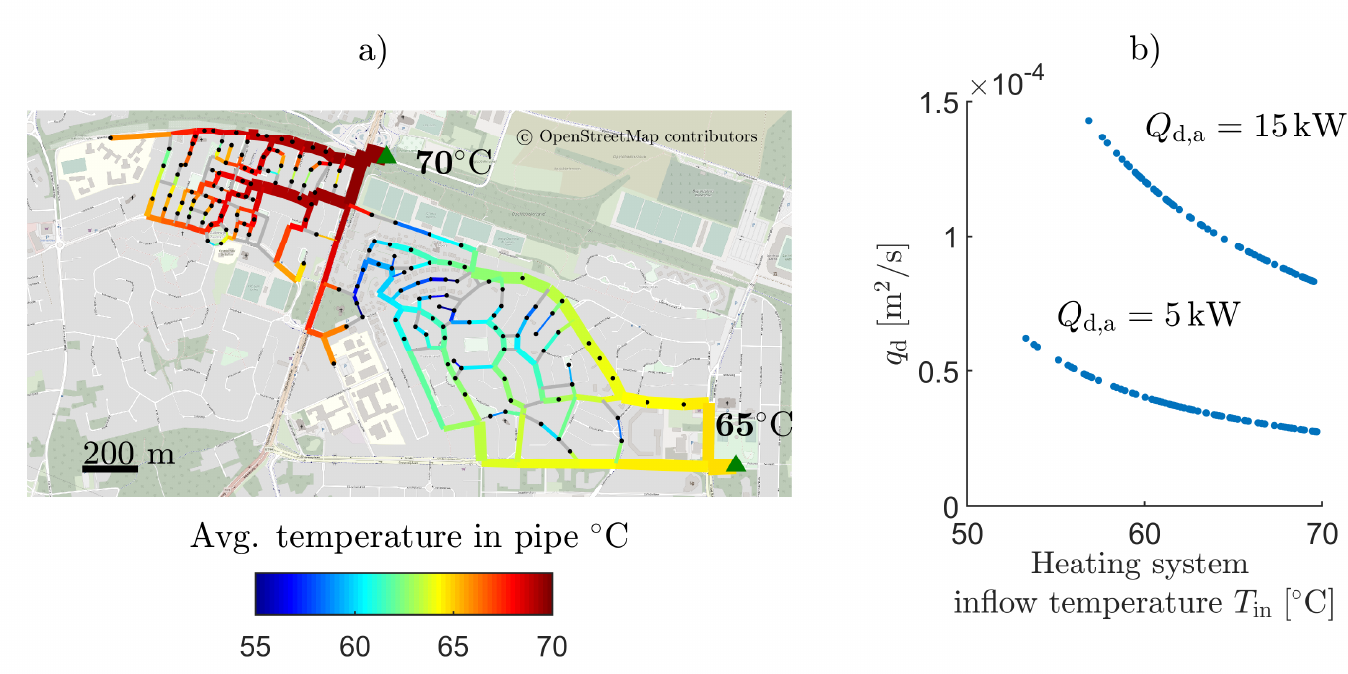}}
	\caption[projection]{a) Average pipe temperature throughout the network. A clear separation in a low temperature network and a high temperature network can be observed. b) Temperature dependency of the flow rate required by the consumers heating system to satisfy the heat demand. In this figure only `dwellings' and `renovated dwellings' are considered. }%
	\label{fig:temperature}%
\end{figure}

It can be seen in \fref{fig:temperature}a that the network is split into a low temperature- and a high temperature network. The temperature differences throughout the network illustrate the importance of basing the network design on a fully-fledged non-linear model for flow and energy modelling. In order to satisfy the heat demand of consumers in the colder parts of the network, the flow rate through the consumers heating system has to be increased (as can be seen in \fref{fig:temperature}b). The increase of flow rate in this test case reaches up to $72\,\%$ when comparing the flow rate for the highest inflow temperatures of $T_{\mathrm{in}}=\SI{69.5}{\degreeCelsius}$ to the lowest inflow temperatures of $T_{\mathrm{in}}=\SI{56.9}{\degreeCelsius}$, for the consumer type `dwelling'. It should be noted that this increase in flow rate influences the cost balance between investment and pump operation cost, ultimately driving the pipe design towards bigger diameters.

\section{Discussion}
We presented a methodology for the automated design of district heating network topologies on the basis of a nonlinear model of flow and heat transport. Because the computational cost of the adjoint sensitivity calculations are intrinsically independent of the number of considered design variables, the methodology is expected to scale well to large heating networks. 

This has been showcased with the design of a fictitious district heating network with a scale well beyond any other applications of nonlinear topological design tools.
On a fictitious design problem of a medium-sized district in Waterschei, the method shows to correctly retrieve a discrete network design that is able to reduce the piping investment by 23\% and the cost related to installed pump capacity and operation by a factor of 14 compared to the initial uniform pipe distribution. While further extensions of the cost function towards the direct optimization of the net present value could still improve the relevancy of the resulting designs, they can easily be incorporated in the current methodology without impacting the computational cost.  

The computational cost in this proof-of-concept design remained well below an hour on a standard laptop. The aggregation of additional design constraints is an important step to achieve this. In the design problem this approach was tested for a constraint that enforces the heat delivered at each single consumer to deviate less than $5 \,\%$ from its heat demand for worst-case conditions. The constraint aggregation method successfully reduced them to a single constraint, while the delivered consumer heat is found to deviate less than $0.008 \,\%$ from the solution that is expected if imposed in a non-aggregated way.

The added value of considering a full nonlinear model in the design clearly follows from the observation that the consumer flows vary up to $72 \,\%$, only because of temperature variations throughout the network. These temperature-induced flow variations in turn influence the optimal pipe design and operational parameters, and are thus important to capture already in the design phase. Linearized models, in contrast, fail to describe the flow pattern in networks with multiple producers or with loops, as well as the heat losses throughout the network and the temperature-dependent consumer heat transfer effectiveness. Nonlinear model-based design is therefore expected to play an increasingly-important role in the design of next-generation networks.

Now that the validity of the method has been established, an important next step is the detailed benchmark of the novel method and its scaling to that of other mixed-integer non-linear programming methods. It is important to note that gradient-based optimization methods like the one presented in this paper could be more sensitive to find local optima than other mixed-integer non-linear programming methods. It should therefore be thoroughly investigated for a number of practical design cases how good the resulting designs perform with respect to those resulting from other design approaches. Finally, it should be mentioned that several uncertainties play a crucial role in the design of thermal networks, such as producer unavailabilites, possible future network extensions, or potential pipe failures. To be of truly practical value, optimal design methods for district heating networks should eventually account for these uncertainties.

\section{Conclusions and outlook}

In this paper, we propose a scalable adjoint-based optimization strategy to determine the optimal network topology, discrete pipe diameters, and operational parameters on the basis of a fully-fledged non-linear flow and energy transport model. Building on recently introduced adjoint approaches for thermal network design with continuous design variables, we introduce a numerical continuation strategy that combines a smoothed projection of the diameters with a penalization of intermediate diameter values to gradually converge the optimization algorithm towards discrete diameter values. As such, an adjoint-based optimal design of network topology and discrete pipe diameters is enabled for the first time.

The scalability of the method is demonstrated with the design of the --to the best of the authors' knowledge-- largest topological network design application reported thus far on the basis of a non-linear transport model. This is enabled here through the intrinsic design-independent scaling of the adjoint method if combined with the suggested constraint aggregation approach. Moreover, by building the optimization on the basis of a nonlinear physics model, heat losses and the temperature-dependent effectiveness of consumer heating systems can be dealt with already in the design phase. The strong variations in local temperatures and flow rates that are found in the test case, confirm the potential impact of these nonlinearities on the optimal design of next-generation thermal networks. 

Future research should aim at setting up a thorough benchmark with standard mixed-integer nonlinear programming tools and developing a methodology to account for uncertainties in the operation conditions.

\section*{Acknowledgements}
Maarten Blommaert is a postdoctoral research fellow of the Research Foundation - Flanders (FWO) and the Flemish Institute for Technological Research (VITO), and Yannick Wack is a doctoral research fellow of VITO.

\appendix

\section{Thermal network equations in vector form} \label{sec:A1}

	To describe the discrete implementation of the solver for both model equations and adjoint equations, we first need to reformulate the equations (1-12) in vector form.	Consider a graph with $n$ nodes and $m$ directed arcs. Let us start by defining the following vectors for flow rates, pressures, temperatures and consumer heat fluxes:
	\begin{equation}
	\begin{aligned}
	&\p = p_i, \ i \in N,  \\
	&\qv = q_{ij}, \ ij \in A, \\
	&\T = \left(\begin{aligned}
	\T_n\\ \T_a
	\end{aligned}\right), \quad \textrm{with} \quad \T_n = \theta_i, \ i \in N \quad\textrm{and}\quad \T_a = \theta_{ij}, \ ij \in A,\\
	&\Q = Q_{ij}, \ ij \in A_{ch}.
	\end{aligned}
	\end{equation}
	Using the same subscripts as used before for the subsets, we introduce vectors related to only one of the subsets $A_i, A_c, A_{ch},A_{cb}, A_p$ of the arcs $A$ in a similar way, e.g. $\qv[i] = q_{ij}, \ ij \in A_i $ denotes the flow in the internal arcs.
	Next, we introduce the node-pipe \emph{incidence matrix}
	\begin{equation*}
	\mtrx{A} = \left(\v{a}_1,\ldots,\v{a}_m\right) \in \R^{n\times m}
	\end{equation*} that contains information on which nodes are connected by a directed arc. That is, each column $\v{a}_i$ contains ``1'' at the start node $i$ of arc $ij$, and a ``-1'' at the end node $j$.
	The continuity condition in the pipe junctions \eqref{eq:con} is then easily translated to vector form as $\mtrx{A}\qv = 0$. The pressure drop over all arcs on the other hand is given by $\v{\Delta}\p = \mtrx{A}^\tp \p$.
	Also here we can easily extend the definition to incidence matrices that include only a number of rows related to a specific subset of arcs, e.g. $\mtrx{A_i} = \left(\mtrx{A}\right)_{i,j}, \ i \in N \  \textrm{and} \ j \in A_i$.

	Using this notation, we can formulate the state equations related to the hydraulic flow succinctly in vector form as $\H(\cv,\y) = 0$, with $\y = \icol{\p\\\qv}$ the hydraulic state vector and  $\cv$ the vector of \emph{design variables} (defined further in \sref{sec:methods}). That is, the vector
	\begin{equation}\label{eq:H}
	\H(\cv,\y) =
	\left(
	\begin{aligned}
	&\mtrx{A} \qv                        \\
	&\mtrx{A_i}^\tp \p - \mtrx{R_i}(\cv,\qv)\qv[i]\\
	&\mtrx{A_c}^\tp \p - \mtrx{R_c}(\cv,\qv)\qv[c]
	\end{aligned}
	\right) = 0,
	\end{equation}
	then contains the residuals of node continuity, pipe momentum, and consumer arc momentum balances, respectively. The diagonal matrices $\mtrx{R_i}(\cv,\qv)$ and $\mtrx{R_c}(\cv,\qv)$ hereby represent the nonlinear flow resistance for internal and consumer arcs, respectively, as can easily be extracted from relations \eqref{eq:1}, \eqref{eq:7}, and \eqref{eq:8}. Note that we leave out the trivial producer conditions and reference pressure condition for simplicity of notation, although they are in principle an integral part of the vector equation $\H(\cv,\y) = 0$.

	To obtain a similar vector equation for energy conservation, we introduce the arc inflow and outflow matrices
	\begin{equation}\begin{aligned}
	\mtrx{A_\textrm{in}} &= \max\!\left(\mtrx{A}\, \diag{\sign{\qv}},0\right)\\
	\mtrx{A_\textrm{out}} &= \min\!\left(\mtrx{A} \,\diag{\sign{\qv}},0\right),
	\end{aligned}\end{equation}
	with ``$\textrm{sgn}$'' the elementwise \emph{sign} function and ``$\textrm{diag}$'' an operator that maps some argument vector $\v{d}$ with elements $d_i$ to a diagonal matrix $\mtrx{D}$ with diagonal elements $\left(\mtrx{D}\right)_{i,i} = d_i$.
	Energy conservation in the pipe junctions, pipes, consumer heating arcs, and bypass arcs can then be succinctly written as a vector equation\footnote{The operator $(\v{a})^{\circ b}$ denotes the \emph{Hadamard exponential}, which simply represents the pointwise exponent of a vector $\v{a}$ to the power $b$.
	}
	\begin{equation}\label{eq:E}
	\E(\cv, \y,\z) =
	\left(
	\begin{aligned}
	&\diag{\mtrx{A_\textrm{in}}\abs{\qv}}\T_n + \mtrx{A_\textrm{out}}\diag{\abs{\qv}}\T_a\\
	&\mtrx{A_{i,\textrm{in}}}^\tp \mtrx{F}(\cv,\y) \T_n - \T_a  \\
	&\rho c_{p} \diag{\qv[ch]} \left(\mtrx{A_{ch,\textrm{in}}}^\tp \T_n - \T_{a,ch}\right) - \xi\left(\tfrac{{A_{ch,\textrm{in}}}^\tp \T_n + \T_{a,ch}}{2}-\theta_{\textrm{house}}\right)^{\circ n}\\
	&\mtrx{A_{cb,\textrm{in}}}^\tp \T_n - \T_{a,cb}
	\end{aligned}
	\right) = 0,
	\end{equation}
	with $\mtrx{F}(\cv,\y)$ a diagonal matrix that has the fractional reduction of $\theta$ through the pipes on its diagonal, as can be deduced from \eqref{eq:3}.
	
	The hydraulic transport equation $\H(\cv,\y) = 0$ and energy transport equation $\E(\cv,\y,\z) = 0$ together form the \emph{state equations} of the thermal network that can be solved for the hydraulic state $\y$ and thermal state $\z$, as will be further elaborated in the next subsection. In an even more compact form, we can thus represent the thermal network model as the vector equation
	\begin{equation}\label{eq:state}
	\c(\cv,\q) = \left(
	\begin{aligned}
	&\H(\cv, \y)\\
	&\E(\cv, \y,\z)
	\end{aligned}\right) = 0,
	\end{equation}
	with $\q = \icol{\y\\ \z} = \icol{\p\\\qv\\\z}$ its \emph{state variables}.

	\section{Implementation of the state solver}\label{sec:A2}
	To efficiently deal with the nonlinear character of the presented model equations, we employ a Newton scheme to iteratively solve the model equations \eqref{eq:state} for an update of the state variables. That is, in each iteration ``$k$", we obtain an update $\Delta\q[k] = \icol{\Delta\y[k]\\ \Delta\z[k]}$ from
	\begin{equation}
	\left[\begin{matrix}
	\Pd{\H}{\y}(\cv,\y[k]) & 0\\
	\Pd{\E}{\y}(\cv,\y[k],\z[k]) & \Pd{\E}{\z}(\cv,\y[k],\z[k])\\
	\end{matrix}\right]
	\left[\begin{matrix}
	\Delta\y[k]\\
	\Delta\z[k]\\
	\end{matrix}\right] = -
	\left[\begin{matrix}
	\H(\cv,\y[k])\\
	\E(\cv,\y[k],\z[k])\\
	\end{matrix}\right]
	\end{equation}
	Note that while the solution of the hydraulic flow equations $\H(\cv,\y) = 0$ does not depend on the temperatures $\z$, the energy advection in the thermal equations $\E(\cv,\y,\z)$ does give rise to a flow dependency. We can therefore solve $\c(\cv,\q) = 0$ for its solution $\qo = \q(\cv)$ in two steps:
	\begin{enumerate}
		\item Solve the hydraulic equations first iteratively for the hydraulic state $\yo$ using updates $\Delta\y[k]$ from
		\begin{equation}\label{eq:19}
		\Pd{\H}{\y}(\cv,\y[k])\Delta\y[k] =  -\H(\cv,\y[k]).
		\end{equation}
		\item Then plug this solution into the energy equations and solve for the thermal state $\zo$ using
		\begin{equation}\label{eq:20}
		\Pd{\E}{\y}(\cv,\yo,\z[k])\Delta\z[k] =  -\E(\cv,\yo,\z[k]).
		\end{equation}
	\end{enumerate}
	This significantly reduces the size of the linear systems that need to be solved.

	It should be noted that the application of this Newton scheme requires the derivatives $\Pd{\H}{\y}$ and $\Pd{\E}{\z}$. These derivatives can be obtained by analytical linearization of \eqref{eq:H} and \eqref{eq:E} with respect to the state variables. Though this is a somewhat cumbersome work, the approach induces fast second-order convergence. Moreover, the approach is particularly interesting in combination with the adjoint approach, as the model linearization is needed again for the adjoint solver.

	\section{Implementation of the adjoint solver}\label{sec:A3}
	The crucial step in the adjoint gradient calculation is the solution of the adjoint equation \eqref{eq:adjoint}. Let us therefore elaborate on the efficient implementation of the adjoint solver, given the thermo-hydraulic model at hand. Rewriting the adjoint equation in terms of $\H$ and $\E$ reveals a similar block-triangular structure as in the state equations:
	\begin{equation}
	\left[\begin{matrix}
	\left(\Pd{\H}{\y}\right)^\tp & \left(\Pd{\E}{\y}\right)^\tp\\
	0 & \left(\Pd{\E}{\z}\right)^\tp
	\end{matrix}\right]
	\left[\begin{matrix}
	\yA\\
	\zA\\
	\end{matrix}\right] = -
	\left[\begin{matrix}
	\left(\Pd{\J}{\y}\right)^\tp\\
	\left(\Pd{\J}{\z}\right)^\tp
	\end{matrix}\right],
	\end{equation}
	with $\qA = \icol{\yA\\\zA}$.

	Block-Gauss elimination therefore leads to the following two-step approach to solve for the adjoint variables $\yA$ and $\zA$:
	\begin{enumerate}
		\item Solve the adjoint thermal equations first iteratively for the adjoint thermal state $\zA$ from
		\begin{equation}
		\left(\Pd{\E}{\z}\right)^\tp \zA =  -\left(\Pd{\J}{\z}\right)^\tp.
		\end{equation}
		\item Then plug this solution into the adjoint hydraulic equations and solve for the adjoint hydraulic state $\yA$ using
		\begin{equation}
		\left(\Pd{\H}{\y}\right)^\tp\yA =  -\left(\Pd{\J}{\y}\right)^\tp - \left(\Pd{\E}{\y}\right)^\tp\zA.
		\end{equation}
	\end{enumerate}
	The adjoint equations thus show a similar sequential structure as the state equations, though in reversed order. Information on the cost function dependence can be as such thought of as propagating backwards through the model equations. Since these equations follow from linearization around $(\cv,\qo)$, the matrices on the left-hand side of the adjoint equations are obtained by simple transposition of the matrices from the last iteration of the model solver (\eqref{eq:19} and \eqref{eq:20}). Because of the linear nature of the adjoint equations, a complete gradient evaluation can thus be performed in only a fraction of the time needed for a model evaluation, independent of the number of design variables.



    \bibliographystyle{ieeetr}
    \bibliography{./literature}

\end{document}